\documentclass[pdflatex,sn-mathphys-num,iicol]{sn-jnl}

\usepackage{breqn}
\usepackage{graphicx}%
\usepackage{multirow}%
\usepackage{amsmath,amssymb,amsfonts}%
\usepackage{amsthm}%
\usepackage{mathrsfs}%
\usepackage[scr=rsfs]{mathalpha}%
\usepackage[title]{appendix}%
\usepackage{xcolor}%
\usepackage{textcomp}%
\usepackage{manyfoot}%
\usepackage{booktabs}%
\usepackage{algorithm}%
\usepackage{algorithmicx}%
\usepackage{algpseudocode}%
\usepackage{listings}%

\usepackage{tikz,xcolor,hyperref}

\definecolor{lime}{HTML}{A6CE39}
\DeclareRobustCommand{\orcidicon}{%
	\begin{tikzpicture}
	\draw[lime, fill=lime] (0,0) 
	circle [radius=0.16] 
	node[white] {{\fontfamily{qag}\selectfont \tiny ID}};
	\draw[white, fill=white] (-0.0625,0.095) 
	circle [radius=0.007];
	\end{tikzpicture}
	\hspace{-2mm}
}

\foreach \x in {A, ..., Z}{%
	\expandafter\xdef\csname orcid\x\endcsname{\noexpand\href{https://orcid.org/\csname orcidauthor\x\endcsname}{\noexpand\orcidicon}}
}

\begin{document}

\title[Machine learning approach to single-shot multiparameter estimation for the NLSE]{Machine learning approach to single-shot multiparameter estimation for the non-linear Schrödinger equation}

\author[1]{\fnm{Louis} \sur{Rossignol}\orcidA{}}

\author[1]{\fnm{Tangui} \sur{Aladjidi}\orcidB{}}

\author[1]{\fnm{Myrann} \sur{Baker-Rasooli}\orcidC{}}

\author*[1]{\fnm{Quentin} \sur{Glorieux}\orcidD{}}\email{quentin.glorieux@sorbonne-universite.fr}

\affil[1]{\orgdiv{Laboratoire Kastler Brossel, Sorbonne Universit\'{e}, CNRS, \'{E}cole Normale Sup\'{e}rieure - Universit\'{e} PSL, Coll\`{e}ge de France}, \orgaddress{\street{4 place Jussieu}, \city{Paris}, \postcode{75005}, \country{France}}}

\abstract{The non-linear Schrödinger equation (NLSE) is a fundamental model to describe  wave dynamics in non-linear media ranging from optical fibers to Bose–Einstein condensates. 
Accurately estimating its parameters (often strongly correlated) from a single measurement remains a significant challenge.
We address this problem by treating parameter estimation as an inverse problem and training a neural network to invert the NLSE mapping. We combine a fast numerical solver with a machine learning approach based on the ConvNeXt architecture and a multivariate Gaussian negative log-likelihood loss function.
 From single-shot field (density and phase) images, our model estimates three key parameters: the non-linear coefficient $n_2$, the saturation intensity $I_{\text{sat}}$, and the linear absorption coefficient $\alpha$.
Trained on 100,000 simulated images, the model achieves a mean absolute error of 3.22\% on 12,500 unseen test samples, demonstrating strong generalization and close agreement with ground truth values. This approach provides an efficient route for characterizing non-linear systems and has the potential to bridge theoretical modeling and experimental data when realistic noise is incorporated.}




\maketitle

\section{Introduction}\label{sec:intro}

The non-linear Schr\"{o}dinger equation (NLSE) provides an effective framework to describe the dynamics of (1+1)D waves in non-linear media, as well as the (2+1)D evolution of the slowly varying envelope of the electric field in the paraxial approximation \cite{Carusotto_2014}.
Its close connection to the Gross–Pitaevskii equation, describing the temporal evolution of the wave function of a weakly interacting quantum gas \cite{pitaevskii_2016_boseeinstein}, has led to the denomination quantum fluids of light \cite{GLORIEUX2025157}.
These fluids of light are now widely employed to investigate superfluidity \cite{michel_2018_superfluid,fontaine2018observation,piekarski2021measurement}, turbulence \cite{PhysRevA.108.063512,eloy2021experimental}, Bose mixtures \cite{piekarski2025spin}, shockwaves \cite{bienaime2021quantitative}, topological excitations \cite{congy2024topological,azam2022vortex, baker2025observation}, and other phenomena usually associated with quantum gases. NLSE has also been applied to investigate non-linear optical fibers \cite{felice2016studynonlinearschrodingerequation, agrawal2010fiberoptic, agrawal_2019_nonlinear} through phenomena like soliton propagation for optical communications \cite{RevModPhys.68.423, 10.3389/fphy.2022.1044845}.
In this work, we address the following question: from a well-defined set of input parameters and known output states, can we recover the parameters governing the evolution?

We developed a model based on recent advancements in convolutional neural networks (CNNs or ConvNets). 
A deep neural network employs multiple processing layers to learn data representations at different levels of abstraction \cite{LeCun2015}. 
These models build upon early work like the Neocognitron \cite{FUKUSHIMA1982455} and were further improved using backpropagation as a learning algorithm \cite{backpropagation}. 
This architecture is widely used in pattern matching tasks such as image labeling and face recognition \cite{alexnet_hinton}. 
Our interest comes from the fact that they were also successfully used in previous works for inverse problems \cite{pmlr-v145-rudi22a, Adler_2017}. 
This makes CNN based approaches a sensible choice for learning the input-output relation of the non-linear Schr\"{o}dinger equation. 

Given the popularity of these methods, variations of our question have already been addressed in the literature. 
In \cite{wang2021inverseproblemnonlinearschrodinger, wang2023cleverneuralnetworksolving}, NLSE models non-linear optical fibers. 
The authors also use CNNs to solve an optimization problem in which they embed NLSE and try to find their intrinsic parameters given input and output. 
While successful, this approach lies into another class of machine learning models called Physics Informed Neural Networks \cite{DBLP:journals/corr/abs-1711-10561}. 
In this work, our goal is that once trained, the model can be used immediately to infer the values of the input parameters without solving an auxiliary optimization problem as in \cite{DBLP:journals/corr/abs-1711-10561}.

In the early 2010s, CNNs dominated deep learning until the introduction of transformers in Attention Is All You Need \cite{Attention}, which led to Vision Transformers (ViTs) \cite{Vision_transformer}. 
ViTs adapted self-attention to images, capturing relationships between image parts without convolutions and achieving high performance. 
Inspired by this, hybrid models such as ConvNeXt \cite{convnext} revisited CNNs with transformer-inspired design improvements, combining the efficiency and inductive capabilities of convolutions with enhanced performance. 
Our model, implemented using PyTorch \cite{Pytorch}, leverages this backbone combined with parameter correlation mapping to provide a reliable method for characterizing the non-linear Schr\"{o}dinger equation within the quantum fluid of light framework in hot atomic vapor \cite{glorieux2023hot}.

\section{Quantum fluids of light}

Following the quantum fluid of light framework introduced in Section~\ref{sec:intro}, the electric field envelope \(\mathcal{E}\) of a monochromatic laser beam propagating through a Kerr non-linear medium \cite{GLORIEUX2025157, Boyd2023} can be effectively described by the non-linear Schr\"{o}dinger equation:

\begin{align}
i \frac{\partial \mathcal{E}}{\partial z}
   = -\frac{1}{2k_0}\nabla_\perp^2 \mathcal{E}
     - i\frac{\alpha}{2}\mathcal{E}
     - k_0 \frac{n_2^{E}}{n_0}|\mathcal{E}|^2 \mathcal{E}.
     \label{eq:nlse}
\end{align}
where \(\mathcal{E}\) is the electric field envelope, $\lambda$ is the wavelength, the wave number is defined as $k_0 = \frac{2\pi}{\lambda}$, $\alpha$ is the linear absorption coefficient, $\nabla_\perp^2 = \partial_x^2 + \partial_y^2$ and $n_2^{E}$ is the non-linear refractive index.
$n_2^{E}$ can be expressed, in terms of the self-Kerr non-linearity $n_2$, as $n_2^{E} = \frac12 \varepsilon_0 n_0 c \, n_2$.

Eq.~\eqref{eq:nlse} contains two driving constants $n_2$ and $\alpha$, that are intrinsic to the propagating medium. 
In the case of atomic vapor, they depend on temperature and the laser's detuning from the atomic resonance. 
Due to saturation, the loss term $\alpha$ as well as the non-linear index $n_2$ will become dependent on intensity, drastically complicating the behavior of this equation. The saturation intensity $I_{\text{sat}}$ describes this behavior through the saturation factor, which damps the dynamics of the propagation.
To account for the effect of intensity saturation, the refractive index is  rescaled by the saturation parameter $\frac{1}{1 + I/I_{\mathrm{sat}}}$.
This saturation parameter also depends in a complex manner on the physical parameters of the medium as well as the beam waist due to atomic transit in hot vapors \cite{aladjidi2022transiteffectsnonlinearindex}.
We thus have three interdependent, and therefore, correlated parameters that will govern the dynamics of the system.

\begin{figure}[h!t]
\centering
\includegraphics[width=0.99\linewidth]{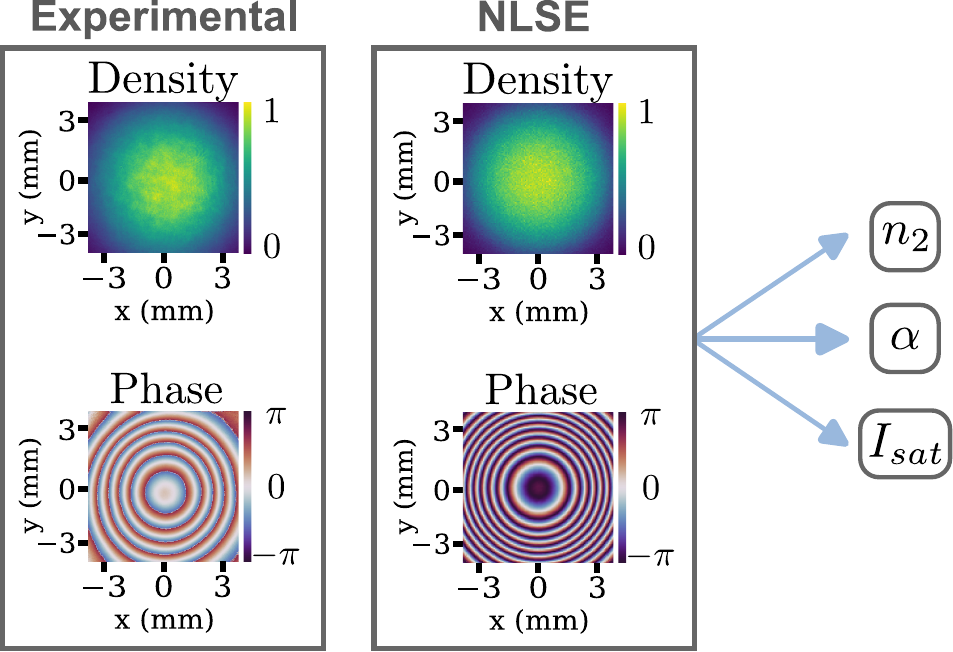}
\caption{\textbf{Data generation and analysis.} –
Typical experimental and numerical density and phase images obtained at the output of the non-linear medium
The data are reproduced with the NLSE package in order to generate multiple datasets for machine learning, enabling the retrieval of $n_2$, $\alpha$, and $I_{\text{sat}}$.}
\label{fig:setup}
\end{figure}

\section{Single shot non-linearity measurement}\label{sec:ML}

\subsection{Training data}
In the laboratory, fluids of light are typically generated by sending a laser beam through a warm rubidium vapor cell ($L$ up to $20$ cm), which serves as a non-linear medium. 
The output field is then recorded on a camera in a single shot using off-axis interferometry and \cite{Cuche:00, aladjidi2022transiteffectsnonlinearindex, GLORIEUX2025157}.
Representative experimental images of the density and phase are shown in Fig.~\ref{fig:setup}, in the top and bottom panels, respectively.
Experimentally, it is challenging to acquire large amounts of data over a wide range of non-linear medium parameters.
For this reason, we numerically simulated the output state in order to generate a large enough dataset to train the model.
The \href{https://github.com/Quantum-Optics-LKB/NLSE}{\texttt{NLSE}} \cite{tanguialadjidi_2024_nlse} Python package provides an efficient and flexible framework to generate multiple training sets and accurately replicates field propagation under laboratory conditions with physical units. 
Inferring the parameters will require a model that can capture the structure of the three-dimensional parameter space  $n_2$, $\alpha$, and $I_{\text{sat}}$. 
To this end, we train our model on a dataset that explores the three-dimensional parameter space for a given experimental setup (laser power, beam size, and non-linear medium length). 
Nonetheless, this method can be generalized for any setup and range because, for reasons we will discuss in the next sections, the model does not embed the physics itself but the ability to map the correlations between the parameters and to infer the three correlated  parameters. 
A different setup will require different optimizer settings, but the rest of the model would remain unchanged.

In this paper we will present results for a laser beam input power $P = 2.1~\text{W}$, a waist $w = 1.7 \times 10^{-3}~\text{m}$ and a propagation length $20 \times 10^{-2}~\text{m}$.
To obtain realistic simulation images, we added Poisson noise to simulate shot noise and Gaussian noise to reproduce the classical thermal noise. 
The generation is repeated for each triplet of parameters: $n_2 \in \left[-1 \times 10^{-9}, -1 \times 10^{-10}\right]~\text{m}^2/\text{W}$, $I_{\text{sat}} \in \left[5 \times 10^4, 1 \times 10^6 \right]~\text{W}/\text{m}^2$, and $\alpha \in \left[1.3 \times 10^1, 3.0 \times 10^1\right]~\text{m}^{-1}$. 
When actually generating the training set, one will have to choose the ranges, the sampling distribution, and the sampling rate, which are critical parameters. 
Here, we generated 50 linearly distributed samples in each range in order to achieve an error that is on the order of the spacing between discrete values.

\begin{figure*}[htbp]
\centering
\includegraphics[width=\textwidth]{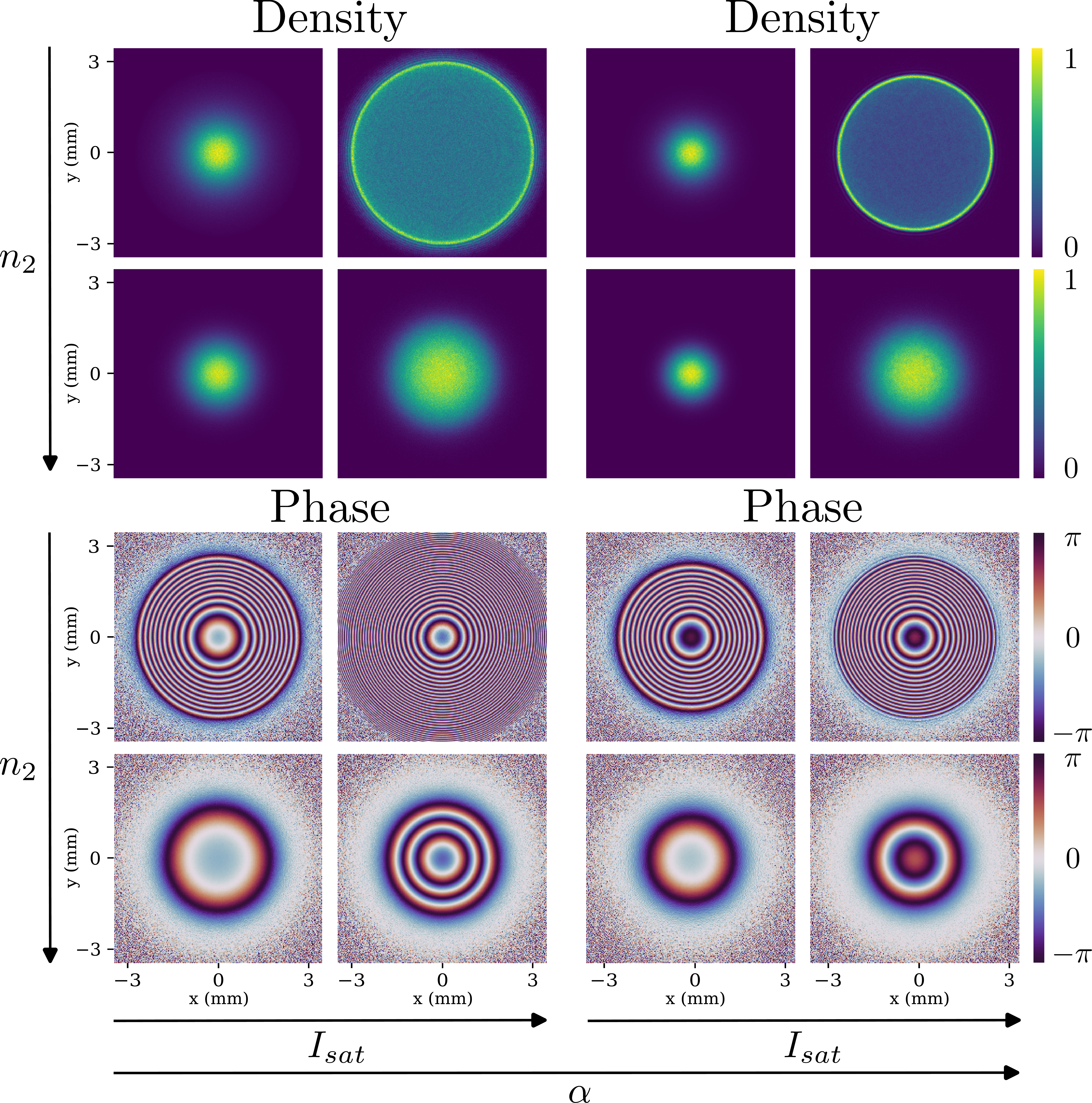}
\caption{\textbf{Parameter space.} — Density and phase profiles used for training, combined as two channels of a single image for each triplet of parameters. 
The simulations correspond to a $780~\text{nm}$ laser beam with input power $P = 2.1~\text{W}$ and waist $w = 1.7 \times 10^{-3}~\text{m}$, propagated through a $20 \times 10^{-2}~\text{m}$ cell.}
\label{fig:density_phase}
\end{figure*}

Fig.~\ref{fig:density_phase} provides a visual representation of the parameter space. 
The images correspond to the NLSE simulations at the output plane of the non-linear medium. 
The top two rows show density profiles, while the bottom two rows display the associated phases. 
Within each group of four images, $n_2$ increases from top to bottom and $I_{\text{sat}}$ from left to right. 
$\alpha$ also increases from left to right.
As we can see on this figure, $n_2$, $\alpha$, and $I_{\text{sat}}$ all contribute to the density profile shape (defocusing) and gaussian non-linear phase across the image. 
However, for each variable, the dynamics occur with different scaling.
Importantly, this figure also highlights that the mapping from parameter space to the space of output profiles is not only bijective but surjective: a profile corresponds to at least one parameter triplet. 
This constitutes the main challenge of this inverse problem, especially if the model must identify the correct triplet from a single observation. 
Having established this, we now turn to the construction of the model. 

\subsection{Architecture}
\begin{figure*}[htbp]
\centering
\includegraphics[width=\textwidth]{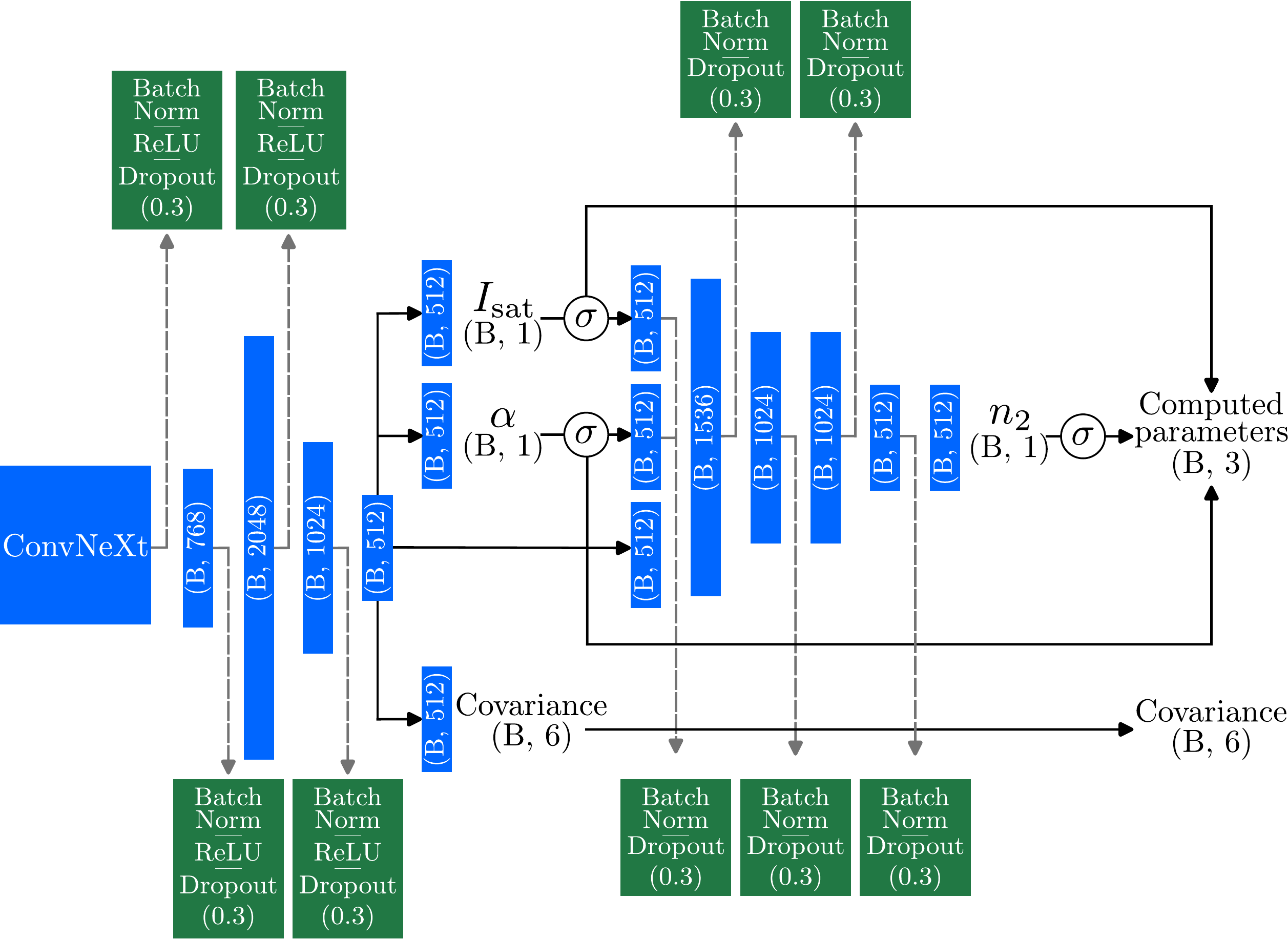}
\caption{\textbf{Schematic of the model architecture.}  –  The blue blocks represent convolutional layers. The green blocks represent the regularization methods applied between layers to improve the gradient propagation. The $\sigma$ represents the sigmoid activation function. The training is done in batches. 
Each batch of our density-phase pair (B, 2, 224, 224) passes through the ConvNeXt backbone. A Feed-Forward neural network is then applied on the output to reduce it to (B, 512). 
We then split the result to be $I_{\text{sat}}$, $\alpha$, and the covariance matrix. This will constitute our pre-prediction. This pre-prediction of $I_{\text{sat}}$ and $\alpha$ is then recombined with the image output inside a new Feed-Forward neural network to then find $n_2$.}
\label{fig:model}
\end{figure*}

As mentioned in the introduction and displayed in Fig.~\ref{fig:model}, we use the Pytorch ConvNeXt architecture as a backbone, which requires input images of size (224px, 224px). 
We feed two channels (density and phase) made of (224, 224) images for each batch $B$ into the network. 

At the output of the ConvNext block, we add a fully-connected neural network (FCNN).
This combination (CNN + FCNN) was shown to significantly improve the performance of CNNs \cite{BASHA2020112}.
However, building neural networks, FCNN in this case, is quite empirical. 
Therefore, the width and depth of the layers might actually be chosen empirically.
Research is striving to try to understand how different structures affect training and how learning is done \cite{fan2024deeprelunetworkssurprisingly, hanin2019complexitylinearregionsdeep}, but it is still not well understood. 
One could, in principle, find a more efficient FCNN and recover the same results by optimizing our model hyperparameters, but this is beyond the scope of the present work.
Nonetheless, we can empirically understand what each block of the model does and what state-of-the-art methods are applied to have a model that performs well. 
To do so, we will design a model to improve \textit{expressivity}. 
In the machine learning literature, this means adding more layers or using larger layers in the model so that it has more weights to be trained on. 
It is known as expressivity since it uses a neural network as a high-dimensional function to approximate a physical system; incrementing the layers simply adds more precision to the model. 
One must, nonetheless, be careful about overfitting, but we address this problem later in this section and in Section \ref{sec:results}. 

With this knowledge, we present the dimensions of the FCNN layers. We go from 768 to 2048 to 1024 to 512 to 1. The sizes were selected empirically: they balance expressivity, training stability, and computational cost, given the size of our dataset and the available GPU memory. 

To mitigate batch-to-batch variability, we normalize the activations \cite{batch_norm} to ensure that no random effects coming from the batches would skew the training. 
Then, we apply the Rectified Linear Unit (ReLU) activation function to improve the gradient propagation \cite{DBLP:journals/corr/abs-1803-08375}, and finish with a dropout layer set to zero out 30$\%$ of each batch. 
This technique is used to improve regularization, reduce overfitting, and prevent co-adaptation of neurons by zeroing a percentage of the data \cite{DBLP:journals/corr/abs-1207-0580}. 
Finally, the output is reduced stepwise to (B,512).

In Section~\ref{sec:reg}, we discuss our choice of loss function (Eq.~\ref{NLL}). 
However, to continue the architecture presentation, we need to mention it. 
Our model predicts the parameters themselves but also the covariance matrix. 
It can be understood that letting the model embed the correlation helps generalization. 
This marks the end of the first stage of the model.

Interestingly, training a model composed of this backbone and the first FCNN has been proved sufficient to estimate $\alpha$ and $I_{\text{sat}}$ but not to recover $n_2$ accurately.
To address this problem, we implement another structure known as conditional neural network.
We first estimate two parameters out of the three in the first FCNN. 
Inside a second FCNN, the predictions are expanded back to dimension (B, 512), again for expressivity, and concatenated with the output of the CNN. 
Finally, we follow a sequence of steps analogous to those of the primary network to infer $n_2$. 
This conditional neural network effectively reduces the space of the parameter search. 
When we first find which $\alpha$ and $I_{\text{sat}}$ match best the data, we evolve in a 2D parameter-space, then when solving for $n_2$ we remain with a 1D $n_2$ parameter-space problem.
A sigmoid activation function is then applied at the output of this secondary model in order to restrict the predictions to the range $[0,1]$.\\

\begin{figure*}[h!t]
\centering
\includegraphics[width=\textwidth]{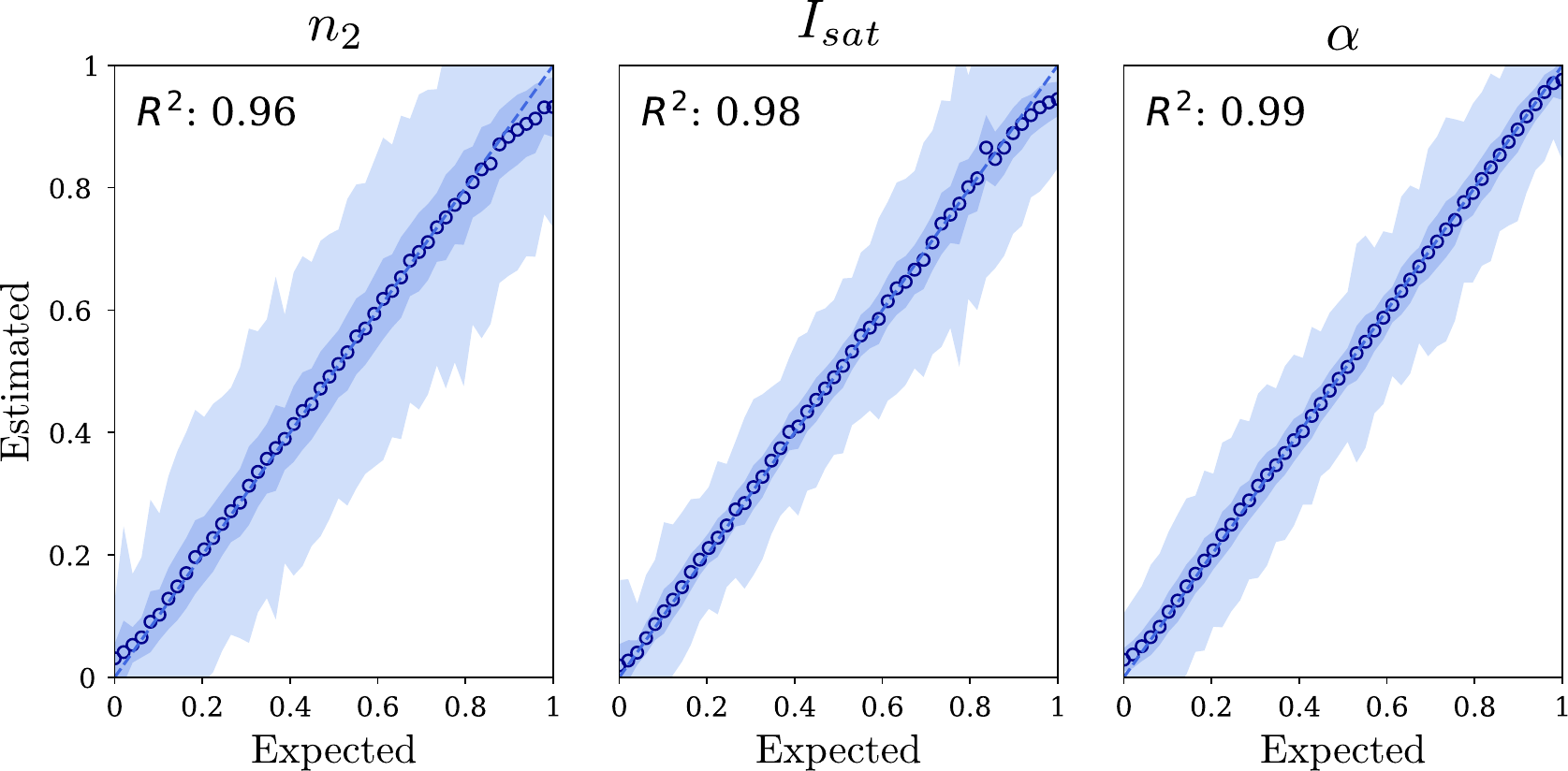}
\caption{\textbf{Comparison of estimated versus expected values.} 
Estimated versus true values for the three parameters: $n_2$ (left), $I_{\text{sat}}$ (center), and $\alpha$ (right), with coefficients of determination $R^2 = 0.96$, $0.98$, and $0.99$, and standard deviations $\sigma = 0.049$, $0.028$, and $0.033$, respectively. 
The dashed line indicates the ideal one-to-one relationship. 
Blue circles represent the trend of the average values, showing slight underfitting near $0$ and overfitting near $1$, less pronounced for $I_{\text{sat}}$ and $\alpha$. 
The dark blue shaded region corresponds to $1\sigma$, while the lighter blue region corresponds to $4\sigma$, illustrating the global regression accuracy.}
\label{fig:predicted_vs_true}
\end{figure*}

\subsection{Regression}\label{sec:reg}

The next key step is to map the model output to the physical parameters of interest. 
Since the parameter triplet forms a continuum in a three-dimensional space, the task is naturally framed as multitask regression. 
A major difficulty arises from the strong correlations between $n_2$, $I_{\text{sat}}$, and $\alpha$: varying one parameter inevitably influences the others. 
As a consequence, standard loss functions such as mean squared error (MSE) or mean absolute error (MAE) are inadequate, as they are univariate. 
To overcome this limitation, we employ a loss function specifically tailored to correlated regression problems: the multivariate Gaussian negative log-likelihood \cite{10.5555/1162264}. 
This approach explicitly accounts for the interdependence among the predicted parameters ($n_2$, $I_{\text{sat}}$, and $\alpha$), and is given by Eq.~\ref{NLL}.
\begin{dmath}
   \mathcal{L} = \frac{1}{2} (\mathbf{x} - \mu)^\top \Sigma^{-1} (\mathbf{x} - \mu) + \frac{1}{2} \log|\Sigma| + \frac{d}{2} \log(2\pi), \label{NLL} 
\end{dmath}
where $\mathbf{x}$ denotes the true parameters, $\mu$ the predicted parameters, and $\Sigma$ the covariance matrix encoding their correlations. 
The covariance matrix characterizes how variations in one parameter relate to variations in the others.
The first term of the loss function, the Mahalanobis distance \cite{mahalanobis}, quantifies the prediction error while explicitly accounting for correlations between parameters. 
The second term, the log-determinant of $\Sigma$, regularizes the predictions by penalizing overconfident uncertainty estimates. 
The third term provides normalization with respect to the dimensionality $d$, here three.

Finally, the regression training is done using the PyTorch Adam-W optimizer. 
This optimizer is based on the widely used Adam optimizer \cite{DBLP:journals/corr/KingmaB14} but modified with a decoupled weight decay \cite{DBLP:journals/corr/abs-1711-05101}, which was shown to help the generalization with Adam. 
We further incorporated two scheduling strategies: reducing the learning rate upon reaching a plateau (i.e., when the loss ceases to decrease), and decreasing the batch size to allow for finer gradient updates during training.

\section{Results}\label{sec:results}

Using the NLSE solver, we generated 50 values each for $n_2$ ($-1 \times 10^{-9}$ to $-1 \times 10^{-10}~\text{m}^2 \cdot \text{W}^{-1}$), $I_{\text{sat}}$ ($5 \times 10^4$ to $1 \times 10^6~\text{W} \cdot \text{m}^{-2}$), and $\alpha$ (13 to 30~$\text{m}^{-1}$).
Using the same input beam parameters as for the training, this procedure yields a dataset of 125,000 samples, each consisting of two-channel images (density and phase) with a resolution of $224 \times 224$ pixels, where $224$ corresponds to the optimized input size for our model. 
The dataset is partitioned into 80\% training, 10\% validation, and 10\% test sets. 
Training is performed for up to 200 epochs with early stopping at epoch 108, using a learning rate of $10^{-4}$ and an effective batch size of 4096.
These training hyperparameters will need fine-tuning when the setup or dataset is changed.
The data generation and training is carried out on an NVIDIA RTX4090 GPU in less than one day. 

After generating the dataset and training the model, we evaluate its performance on the test set, which was not seen during training, to assess its generalization to unseen data. 
The primary evaluation metric is the MAE  across all parameters, which is measured at 3.22 \%. 
Additionally, the coefficient of determination, $R^2 = 0.977$, quantifies the goodness of fit, with values close to 1 indicating accurate predictions. 
This metric reflects the discrepancy between the predicted and true values for the three parameters. 
Specifically, $n_2$ is predicted with a MAE of 4.28 \%, $I_{\text{sat}}$ with 2.87 \%, and $\alpha$ with 2.51 \%. 
The model performance is shown in the predicted-versus-true plots shown in Fig.~\ref{fig:predicted_vs_true}.
The MAE results indicate a slightly larger discrepancy for $n_2$, which exhibits a sparser distribution, compared to $I_{\text{sat}}$ and $\alpha$. 
Overall, the model demonstrates strong predictive performance, with standard deviations around 0.025 and $R^2$ values close to 1. 
Although $n_2$ is predicted with slightly lower accuracy than the other parameters, the model still provides a reliable estimate of the evolution parameters.

\section{Conclusion}

In this work, we developed a method to determine the parameters $(n_2, I_{\text{sat}}, \alpha)$, characterizing the evolution of a paraxial fluid of light in a warm vapor. 
To generate training data matching experimental observations, we employed the \texttt{NLSE} Python library and designed a machine learning architecture leveraging recent advances in convolutional neural networks and multivariate loss functions. 
We demonstrate that the model accurately maps the simulated images to the corresponding parameters with low error rates. 
For future work, it will be possible to implement a refined mapping by incorporating alternative noise sources to better emulate experimental conditions and apply the method to experimental data. 
Overall, our study suggests that machine-learning approaches could be valuable across a wide range of physical systems obeying NLSE-like non-linear dynamics.

\backmatter


\bmhead{Acknowledgements}

The authors warmly thank C. Piekarski, Q. Schibler, and T. Bourgeois for highly valuable discussions that helped the development of the model and the understanding of the physical system.

\begin{itemize}
\item Code availability: the code is further explained in detail and shared on the GitHub of the project named \href{https://github.com/Quantum-Optics-LKB/xp2nlse-ml}{XP2NLSE-ML}.
\item Author contribution:
LR developed the model and its Python implementation 
TA developed \cite{tanguialadjidi_2024_nlse} and participated in the discussion on the implementation of the model.
MBR provided experimental data and feedback on the method.
QG provided the general supervision of the project.
All authors participated to write the manuscript.
\end{itemize}

\begin{appendices}




\end{appendices}
\bibliography{main}

\end{document}